\documentclass[11pt]{revtex4}    
\usepackage{amssymb,epsf}
\usepackage{latexsym}
\begin{document}
\title{Thermodynamics of interacting holographic dark energy\\ with apparent horizon as an IR cutoff}
\author{Ahmad Sheykhi \footnote{
sheykhi@mail.uk.ac.ir}}
\address{Department of Physics, Shahid Bahonar University, P.O. Box 76175, Kerman, Iran\\
         Research Institute for Astronomy and Astrophysics of Maragha (RIAAM), Maragha,
         Iran}

 \begin{abstract}
As soon as an interaction between holographic dark energy and dark
matter is taken into account, the identification of IR cutoff with
Hubble radius $H^{-1}$, in flat universe, can simultaneously drive
accelerated expansion and solve the coincidence problem. Based on
this, we demonstrate that in a non-flat universe the natural
choice for IR cutoff could be the apparent horizon radius,
$\tilde{r}_A={1}/{\sqrt{H^2+k/a^2}}$. We show that any interaction
of dark matter with holographic dark energy, whose infrared cutoff
is set by the apparent horizon radius, implies an accelerated
expansion and a constant ratio of the energy densities of both
components thus solving the coincidence problem. We also verify
that for a universe filled with dark energy and dark matter the
Friedmann equation can be written in the form of the modified
first law of thermodynamics, $dE=T_hdS_h+WdV$, at apparent
horizon. In addition, the generalized second law of thermodynamics
is fulfilled in a region enclosed by the apparent horizon. These
results hold regardless of the specific form of dark energy and
interaction term. Our study might reveal that in an accelerating
universe with spatial curvature, the apparent horizon is a
physical boundary from the thermodynamical point of view.
\end{abstract}

 \maketitle
  \section{Introduction\label{Intro}}
The combined analysis of cosmological observations reveal that
nearly three quarters of our universe consists of a mysterious
energy component usually dubbed ``dark energy'' which is
responsible for the cosmic expansion, and the remaining part
consists of pressureless matter  \cite{Rie}. The nature of such
previously unforeseen energy still remains a complete mystery,
except for the fact that it has negative pressure. In this new
conceptual set up, one of the important questions concerns the
thermodynamical behavior of the accelerated expanding universe
driven by dark energy. It is important to ask whether
thermodynamics in an accelerating universe can reveal some
properties of dark energy. The profound connection between
thermodynamics and gravity has been observed in the cosmological
situations
\cite{Cai2,Cai3,CaiKim,Fro,Wang,Cai4,Shey1,Shey2,Shey3}. This
connection implies that the thermodynamical properties can help
understand the dark energy, which gives strong motivation to study
thermodynamics in the accelerating universe. It is also of great
interest to investigate the validity of the generalized second law
of thermodynamics in the accelerating universe driven by dark
energy \cite{wangb0}. The generalized second law of thermodynamics
is an important principle in governing the development of the
nature.

An interesting attempt for probing the nature of dark energy
within the framework of quantum gravity, is the so-called
``Holographic Dark Energy" (HDE) proposal. This model which has
arisen a lot of enthusiasm recently
\cite{Coh,Li,Huang,Hsu,HDE,Setare1,wang0,wang1}, is motivated from
the holographic hypothesis \cite{Suss1} and has been tested and
constrained by various astronomical observations \cite{Xin}. It is
important to note that in the literature, various scenarios of HDE
have been studied via considering different system's IR cutoff. In
the absence of interaction between dark matter and dark energy in
flat universe, Li \cite{Li} discussed three choices for the length
scale $L$ which is supposed to provide an IR cutoff. The first
choice is the Hubble radius, $L=H^{-1}$ \cite{Hsu}, which leads to
a wrong equation of state, namely that for dust. The second option
is the particle horizon radius. In this case it is impossible to
obtain an accelerated expansion. Only the third choice, the
identification of $L$ with the radius of the future event horizon
gives the desired result, namely a sufficiently negative equation
of state to obtain an accelerated universe.

However, as soon as an interaction between dark energy and dark
matter is taken into account, the first choice, $L=H^{-1}$, in
flat universe, can simultaneously drive accelerated expansion and
solve the coincidence problem \cite{pav1}. Based on this, we
demonstrate that in a non-flat universe the natural choice for IR
cutoff could be the apparent horizon radius. We show that any
interaction of pressureless dark matter with HDE, whose infrared
cutoff is set by the apparent horizon radius, implies a constant
ratio of the energy densities of both components thus solving the
coincidence problem. Besides, it was argued that for an
accelerating universe inside the event horizon the generalized
second law does not satisfy, while the accelerating universe
enveloped by the Hubble horizon satisfies the generalized second
law \cite{Jia}. This implies that the event horizon in an
accelerating universe might not be a physical boundary from the
thermodynamical point of view. Thus, it looks that we need to
define a convenient horizon that satisfies all of our accepted
principles in a universe with any spacial curvature. In the next
section, we study the interacting HDE with apparent horizon as an
IR cutoff. In section \ref{FIRST}, we examine the first law of
thermodynamics on the apparent horizon in an accelerating universe
with spacial curvature. In section \ref{GSL}, we investigate the
validity of the generalized second law of thermodynamics in  a
region enclosed by the apparent horizon. The last section is
devoted to conclusions.

 \section{Interacting HDE with apparent horizon as an IR cutoff\label{IR}}
We consider a homogenous and isotropic Friedmann-Robertson-Walker
(FRW) universe which is described by the line element
\begin{equation}
ds^2={h}_{\mu \nu}dx^{\mu}
dx^{\nu}+\tilde{r}^2(d\theta^2+\sin^2\theta d\phi^2),
\end{equation}
where $\tilde{r}=a(t)r$, $x^0=t, x^1=r$, the two dimensional
metric $h_{\mu \nu}$=diag $(-1, a^2/(1-kr^2))$. Here $k$ denotes
the curvature of space with $k = 0, 1, -1$ corresponding to open,
flat, and closed universes, respectively. A closed universe with a
small positive curvature ($\Omega_k\simeq0.01$) is compatible with
observations \cite{spe}. Then, the dynamical apparent horizon, a
marginally trapped surface with vanishing expansion, is determined
by the relation $h^{\mu \nu}\partial_{\mu}\tilde
{r}\partial_{\nu}\tilde {r}=0$, which implies that the vector
$\nabla \tilde {r}$ is null on the apparent horizon surface. The
apparent horizon was argued as a causal horizon for a dynamical
spacetime and is associated with gravitational entropy and surface
gravity \cite{Hay2,Bak}. A simple calculation gives the apparent
horizon radius for the FRW universe
\begin{equation}
\label{radius}
 \tilde{r}_A=\frac{1}{\sqrt{H^2+k/a^2}}.
\end{equation}
The corresponding Friedmann equation takes the form
\begin{eqnarray}\label{Fried}
H^2+\frac{k}{a^2}=\frac{8\pi G}{3} \left( \rho_m+\rho_D \right),
\end{eqnarray}
where $\rho_m$ and $\rho_D$ are the energy density of dark matter
and dark energy inside apparent horizon, respectively. Since we
consider the interaction between dark matter and dark energy,
$\rho_{m}$ and $\rho_{D}$ do not conserve separately; they must
rather enter the energy balances
\begin{eqnarray}
&&\dot{\rho}_m+3H\rho_m=Q, \label{consm}
\\&& \dot{\rho}_D+3H\rho_D(1+w_D)=-Q.\label{consq}
\end{eqnarray}
where $w_{D}=p_D/\rho_D$ is the equation of state parameter of
HDE,  and $Q$ stands for the interaction term. We also ignore the
baryonic matter ($\Omega_{BM}\approx0.04$) in comparison with dark
matter and dark energy ($\Omega_{DM}+\Omega_{DE}\approx0.96$). We
shall assume the ansatz $Q =\Gamma\rho_D$ with $\Gamma>0$ which
means that there is an energy transfer from the dark energy to
dark matter. It is important to note that the continuity equations
imply that the interaction term should be a function of a quantity
with units of inverse of time (a first and natural choice can be
the Hubble factor $H$) multiplied with the energy density.
Therefore, the interaction term could be in any of the following
forms: (i) $Q\propto H \rho_D$, (ii) $Q\propto H
\rho_m$, or (iii) $Q\propto H (\rho_m+\rho_D)$. However, we can present the above three forms in one
expression as $Q =\Gamma\rho_D$, where
\begin{eqnarray}
\begin{array}{ll}
\Gamma=3b^2H  \hspace{1.3cm}   {\rm for}\  \  Q\propto H \rho_D, &  \\
\Gamma=3b^2Hu \hspace{1.1cm}   {\rm for} \  \ Q\propto H \rho_m,&  \\
\Gamma=3b^2H(1+u) \ \   {\rm for} \  \ Q\propto H
(\rho_m+\rho_D),&
  \end{array}
 \end{eqnarray}
with $b^2$ is a coupling constant and $u =\rho_m/\rho_D$ is the
ratio of energy densities. The freedom of choosing the specific
form of the interaction term $Q$ stems from our incognizance of
the origin and nature of dark energy as well as dark matter.
Moreover, a microphysical model describing the interaction between
the dark components of the universe is not available nowadays. If
we introduce, as usual, the fractional energy densities such as
\begin{eqnarray}\label{Omega}
\Omega_m=\frac{8\pi G\rho_m}{3H^2}, \hspace{0.5cm}
\Omega_D=\frac{8\pi G\rho_D}{3H^2},\hspace{0.5cm}
\Omega_k=\frac{k}{H^2 a^2},
\end{eqnarray}
then, the Friedmann equation can be written as
$\Omega_m+\Omega_D=1+\Omega_k$. In terms of the apparent horizon
radius, we can rewrite the Friedmann equation as
\begin{equation}
\label{Fri2}
 \frac{1}{\tilde {r}_{A}^2}=\frac{8\pi G}{3}\left(\rho_m+\rho_D\right).
 \end{equation}
For completeness, we give the deceleration parameter
\begin{equation}\label{q}
 q=-\frac{\ddot{a}}{a H^2}=-1-\frac{\dot{H}}{H^2},
 \end{equation}
 which combined with the Hubble parameter and the dimensionless
density parameters form a set of useful parameters for the
description of the astrophysical observations. It is a matter of
calculation to show that
\begin{equation}\label{q2}
 q=-(1+\Omega_k)+\frac{3}{2}\Omega_D(1+u+w_D).
  \end{equation}
The evolution of $u$ is governed by
\begin{equation}\label{udot}
 \dot{u}=3Hu\left[w_D+\frac{1+u}{u}\frac{\Gamma}{3H}\right].
 \end{equation}
We assume the HDE density has the form
\begin{equation}\label{rhoD}
 \rho_D=\frac{3c^2}{8\pi G\tilde
 {r}_{A}^{2}},
 \end{equation}
where $c^2$ is a constant, the coefficient $3$ is for convenient,
and we have set the apparent horizon radius $L={\tilde {r}_{A}}$
as system's IR cutoff in holographic model of dark energy.
Inserting Eq. (\ref{rhoD}) in Eq. (\ref{Fri2}) immediately yields
\begin{equation}
 \rho_m=\frac{3(1-c^2)}{8\pi G\tilde
 {r}_{A}^{2}}.
\end{equation}
Thus we reach
\begin{equation}
 u=\frac{\rho_m}{\rho_D}=\frac{1-c^2}{c^2}.
\end{equation}
This implies that the ratio of the energy densities is a constant;
thus the coincidence problem can be solved. Taking the derivative
of Eq. (\ref{rhoD}) we get
\begin{equation}\label{rhoD2}
 \dot{\rho}_D=-2\rho_D\frac{\dot{\tilde
 {r}_{A}}}{\tilde
 {r}_{A}}=-3c^2 H \rho_D(1+u+w_D).
 \end{equation}
 where we have employed Eqs. (\ref{consm}), (\ref{consq}) and (\ref{Fri2}). Combining this
 equation with (\ref{consq}) we obtain
\begin{equation}\label{wD}
 w_D=-\left(1+\frac{1}{u}\right)\frac{\Gamma}{3H}.
 \end{equation}
 Substituting $w_D$ into (\ref{q2}), we find
\begin{equation}\label{q32}
 q=-(1+\Omega_k)-\frac{3}{2}\Omega_D(1+u)\left(\frac{\Gamma}{3Hu}-1\right).
  \end{equation}
The interaction parameter $\frac{\Gamma}{3H}$ together with the
energy density ratio $u$ determine the equation of state
parameter. In the absence of interaction, we encounter dust with
$w_D =0$. For the choice $L=\tilde {r}_{A}$ an interaction is the
only way to have an equation of state different from that for
dust. Any decay of the dark energy component into pressureless
matter is necessarily accompanied by an equation of state $w_D <
0$. The existence of an interaction has another interesting
consequence. Inserting expression $w_D$ into (\ref{udot}) leads to
$\dot{u} = 0$, i.e., $u$ = const. Thus, any interaction of dark
matter with HDE, whose infrared cutoff is set by the apparent
horizon radius, implies an accelerated expansion and a constant
ratio of the energy densities, irrespective of the specific
structure of the interaction. It is important to note that
although choosing $L=H^{-1}$, in a spatially flat universe, can
drive accelerated expansion and solve the coincidence problem
\cite{pav1}, but taking into account the spatial curvature term
gives rise to an additional dynamics which implies a small
(compared with the Hubble rate) change of the energy density
ratio; thus the coincidence problem cannot be solved exactly (see
\cite{pav2} for details). This implies that in an accelerating
universe with spacial curvature the Hubble radius $H^{-1}$ is not
a convenient choice.

In summary, in a universe with spacial curvature, the
identification of IR cutoff with apparent horizon radius $\tilde
{r}_{A}$ is not only the most obvious but also the simplest choice
which can simultaneously drive accelerated expansion and solve the
coincidence problem. It is important to note that the interaction
is essential to simultaneously solve the coincidence problem and
have late acceleration. There is no non-interacting limit, since
in the absence of interaction, i.e., $\Gamma=0$, there is no
acceleration.

\section{First law of thermodynamics\label{FIRST}}
In this section we are going to examine the first law of
thermodynamics. In particular, we show that for a closed universe
filled with HDE and dark matter the Friedmann equation can be
written directly in the form of the modified first law of
thermodynamics at apparent horizon regardless of the specific form
of the dark energy. The associated temperature with the apparent
horizon can be defined as $T = \kappa/2\pi$, where $\kappa$ is the
surface gravity $
 \kappa =\frac{1}{\sqrt{-h}}\partial_{\mu}\left(\sqrt{-h}h^{\mu \nu}\partial_{\mu\nu}\tilde
 {r}\right).
$ Then one can easily show that the surface gravity at the
apparent horizon of FRW universe can be written as
\begin{equation}\label{surgrav} \kappa=-\frac{1}{\tilde
r_A}\left(1-\frac{\dot {\tilde r}_A}{2H\tilde r_A}\right).
\end{equation}
When $\dot {\tilde r}_A\leq 2H\tilde r_A$, the surface gravity
$\kappa\leq 0$, which leads the temperature $T\leq 0$ if one
defines the temperature of the apparent horizon as $T=\kappa/2\pi$
. Physically it is not easy to accept the negative temperature,
the temperature on the apparent horizon should be defined as
$T=|\kappa|/2\pi$. Recently the connection between temperature on
the apparent horizon and the Hawking radiation has been considered
in \cite{cao}, which gives more solid physical implication of the
temperature associated with the apparent horizon.

Taking differential form of equation (\ref{Fri2}) and using Eqs.
(\ref{consm}) and (\ref{consq}), we can get the differential form
of the Friedmann equation
\begin{equation} \label{Frid3}
\frac{1}{4\pi G} \frac{d\tilde {r}_{A}}{\tilde
{r}_{A}^3}=H\rho_D\left(1+u+w_{D}\right)dt.
\end{equation}
Multiplying both sides of the equation (\ref{Frid3}) by a factor
$4\pi\tilde{r}_{A}^{3}\left(1-\frac{\dot{\tilde{r}}_{A}}{2H\tilde
r_A}\right)$, and using the expression (\ref{surgrav}) for the
surface gravity, after some simplification one can rewrite this
equation in the form
\begin{eqnarray}
\label{Frid4} -\frac{\kappa}{2\pi }\frac{2\pi\tilde {r}_{A}d\tilde
{r}_{A}}{G}&=&4\pi\tilde
 {r}_{A}^{3}H\rho_D\left(1+u+w_{D}\right)dt-2\pi\tilde
 {r}_{A}^{2}\rho_D\left(1+u+w_{D}\right)d\tilde {r}_{A}.
 \end{eqnarray}
$E=(\rho_m+\rho_D) V$ is the total energy content of the universe
inside a $3$-sphere of radius $\tilde{r}_{A}$, where
$V=\frac{4\pi}{3}\tilde{r}_{A}^{3}$ is the volume enveloped by
3-dimensional sphere with the area of apparent horizon
$A=4\pi\tilde{r}_{A}^{2}$. Taking differential form of the
relation $E=(\rho_m+\rho_D) \frac{4\pi}{3}\tilde{r}_{A}^{3}$ for
the total matter and energy inside the apparent horizon, we get
\begin{equation}
\label{dE1}
 dE=4\pi\tilde
 {r}_{A}^{2}(\rho_m+\rho_D) d\tilde {r}_{A}+\frac{4\pi}{3}\tilde{r}_{A}^{3}(\dot{\rho}_{m}+\dot{\rho}_{D}) dt.
\end{equation}
Using Eqs. (\ref{consm}) and (\ref{consq}), we obtain
\begin{equation}
\label{dE2}
 dE=4\pi\tilde
 {r}_{A}^{2}\rho_D(1+u) d\tilde {r}_{A}-4\pi  \tilde{r}_{A}^{3}H \rho_D\left(1+u+w_{D}\right)dt.
\end{equation}
Substituting this relation into (\ref{Frid4}), and using the
relation between temperature and the surface gravity, we get the
modified first law of thermodynamics on the apparent horizon
\begin{equation}\label{FL}
dE = T_h dS_h + WdV,
\end{equation}
where $S_{h} ={A}/{4G}$ is the entropy associated to the apparent
horizon, and
\begin{equation}W=\frac{1}{2}(\rho_m+\rho_D-p_D)=\frac{1}{2}\rho_D\left(1+u-w_{D}\right)\end{equation}
is the matter work density \cite{Hay2}. The work density term is
regarded as the work done by the change of the apparent horizon,
which is used to replace the negative pressure if compared with
the standard first law of thermodynamics, $dE = TdS-pdV$. For a
pure de Sitter space, $\rho_m+\rho_D=-p_D$, then our work term
reduces to the standard $-p_DdV$ and we obtain exactly the first
law of thermodynamics.
\section{Generalized Second law of thermodynamics\label{GSL}}
In this section we turn to investigate the validity of the
generalized second law of thermodynamics in  a region enclosed by
the apparent horizon. Differentiating Eq. (\ref{Fri2}) with
respect to the cosmic time and using Eqs. (\ref{consm}) and
(\ref{consq}) we get
\begin{equation} \label{dotr1}
\dot{\tilde{r}}_{A}=4\pi G H {\tilde{r}_{A}^3} \rho_D(1+u+w_D).
\end{equation}
One can see from the above equation that $\dot{\tilde{r}}_{A}>0$
provided condition $w_D>-1-u$, holds. Let us now turn to find out
$T_{h} \dot{S_{h}}$:
\begin{equation}\label{TSh}
T_{h} \dot{S_{h}} =\frac{1}{2\pi \tilde r_A}\left(1-\frac{\dot
{\tilde r}_A}{2H\tilde r_A}\right)\frac{d}{dt} \left(\frac{\pi
\tilde{r}_{A}^2 }{G}\right).
\end{equation}
After some simplification and using Eq. (\ref{dotr1}) we get
\begin{equation}\label{TSh1}
T_{h} \dot{S_{h}} =4\pi H {\tilde{r}_{A}^3}
\rho_D(1+u+w_D)\left(1-\frac{\dot {\tilde r}_A}{2H\tilde
r_A}\right).
\end{equation}
As we argued above the term $\left(1-\frac{\dot {\tilde
r}_A}{2H\tilde r_A}\right)$ is positive to ensure $T_{h}>0$,
however, in an accelerating universe the equation of state
parameter of dark energy may cross the phantom divide, i.e.,
$w_D<-1-u$. This indicates that the second law of thermodynamics,
$\dot{S_{h}}\geq0$, does not hold on the apparent horizon. Then
the question arises, ``will the generalized second law of
thermodynamics, $\dot{S_{h}}+\dot{S_{m}}+\dot{S_{D}}\geq0$, can be
satisfied in a region enclosed by the apparent horizon?'' The
entropy of dark energy plus dark matter inside the apparent
horizon, $S=S_{m}+S_{D}$, can be related to the total energy
$E=(\rho_m+\rho_D) V$ and pressure $p_D$ in the horizon by the
Gibbs equation \cite{Pavon2}
\begin{equation}\label{Gib1}
T dS=d[(\rho_m+\rho_D) V]+p_DdV=V(
d\rho_m+d\rho_D)+\rho_D(1+u+w_D)dV,
\end{equation}
where $T=T_{m}=T_D$ and $S=S_{m}+S_D$ are the temperature and the
total entropy of the energy and matter content inside the horizon,
respectively. Here we assumed that the temperature of both dark
components are equal, due to their mutual interaction. We also
limit ourselves to the assumption that the thermal system bounded
by the apparent horizon remains in equilibrium so that the
temperature of the system must be uniform and the same as the
temperature of its boundary. This requires that the temperature
$T$ of the energy content inside the apparent horizon should be in
equilibrium with the temperature $T_h$ associated with the
apparent horizon, so we have $T=T_h$\cite{Pavon2}. This expression
holds in the local equilibrium hypothesis. If the temperature of
the fluid differs much from that of the horizon, there will be
spontaneous heat flow between the horizon and the fluid and the
local equilibrium hypothesis will no longer hold. This is also at
variance with the FRW geometry. In general, when we consider the
thermal equilibrium state of the universe, the temperature of the
universe is associated with the apparent horizon. Therefore from
the Gibbs equation (\ref{Gib1}) we can obtain
\begin{equation}\label{TSm1}
T_{h} (\dot{S_{m}}+\dot{S_{D}}) =4\pi {\tilde{r}_{A}^2}
\rho_D(1+u+w_D)\dot{\tilde{r}}_{A}-4\pi H {\tilde{r}_{A}^3}
\rho_D(1+u+w_D).
\end{equation}
To check the generalized second law of thermodynamics, we have to
examine the evolution of the total entropy $S_h + S_m+S_D$. Adding
equations (\ref{TSh1}) and (\ref{TSm1}),  we get
\begin{equation}\label{GSL1}
T_{h}( \dot{S}_{h}+\dot{S}_{m}+\dot{S}_D)=2\pi {\tilde{r}_{A}^2}
\rho_D(1+u+w_D)\dot{\tilde{r}}_A=\frac{A}{2}\rho_D(1+u+w_D) \dot
{\tilde r}_A.
\end{equation}
where $A>0$ is the area of apparent horizon. Substituting $\dot
{\tilde r}_A$ from Eq. (\ref{dotr1}) into (\ref{GSL1}) we get
\begin{equation}\label{GSL11}
T_{h}( \dot{S}_{h}+\dot{S}_{m}+\dot{S}_D)=2\pi G A H {\tilde
r_A}^{3} \rho^2_D(1+u+w_D)^2.
\end{equation}
The right hand side of the above equation cannot be negative
throughout the history of the universe, which means that $
\dot{S_{h}}+\dot{S_{m}}+\dot{S_{D}}\geq0$ always holds. This
indicates that for a universe with spacial curvature filled with
interacting dark components, the generalized second law of
thermodynamics is fulfilled in a region enclosed by the apparent
horizon.
\section{conculusions\label{Con}}
It is worthwhile to note that in the literature, various scenarios
of HDE have been studied via considering different system's IR
cutoff. In the absence of interaction the convenient choice for
the IR cutoff are the radial size of the horizon $R_h$  and the
radius of the event horizon measured on the sphere of the horizon
$L=ar(t)$ in spatially flat and curved universe, respectively.
Although, in these cases the HDE gives the observation value of
dark energy in the universe and can drive the universe to an
accelerated expansion phase, but an obvious drawback concerning
causality appears. Event horizon is a global concept of spacetime;
existence of event horizon of the universe depends on future
evolution of the universe; and event horizon exists only for
universe with forever accelerated expansion. However, as soon as
an interaction between dark energy and dark matter is taken into
account, the identification of $L$ with $H^{-1}$ in flat universe,
can simultaneously drive accelerated expansion and solve the
coincidence problem \cite{pav1}. The Hubble radius is not only the
most obvious but also the simplest choice in flat universe.

In this paper, we demonstrated that in a universe with spacial
curvature the natural choice for IR cutoff could be the apparent
horizon radius, $\tilde{r}_A={1}/{\sqrt{H^2+k/a^2}}$. We showed
that any interaction of pressureless dark matter with HDE, whose
infrared cutoff is set by the apparent horizon radius, implies a
constant ratio of the energy densities of both dark components
thus solving the coincidence problem. In addition, we examined the
validity of the first and the generalized second law of
thermodynamics for a universe filled with mutual interacting dark
components in a region enclosed by the apparent horizon. These
results hold regardless of the specific form of dark energy and
interaction term $Q$. Our study further supports that in a
universe with spatial curvature, the apparent horizon is a
physical boundary from the thermodynamical point of view.

\acknowledgments{I thank the anonymous referee for constructive
comments. I am also grateful to Prof. B. Wang for helpful
discussions and reading the manuscript. This work has been
supported by Research Institute for Astronomy and Astrophysics of
Maragha.}

\end{document}